\documentclass[aps]{revtex4}
\usepackage{graphicx}
\usepackage{amsmath}

\begin{document}

\title{Spherically Symmetric solutions in Multidimensional 
Gravity with the  SU(2) Gauge Group as the Extra Dimensions} 
\author{V. Dzhunushaliev}
\email{dzhun@hotmail.kg}
\affiliation{Dept. Phys.and Microel. Engineer., Kyrgyz-Russian
Slavic University, Bishkek, Kievskaya Str. 44, 720000, Kyrgyz
Republic}
\author{H.-J. Schmidt}
\email{hjschmi@rz.uni-potsdam.de}
\homepage{http://www.physik.fu-berlin.de/hjschmi}
\affiliation{Institut f\"ur Theoretische Physik, Freie
Universit\"at Berlin\\
and \\
Institut f\"ur Mathematik, Universit\"at Potsdam 
PF 601553, D-14415 Potsdam, Germany}
\author{O. Rurenko}
\affiliation{Dept. Theor. Phys., Kyrgyz State National University, 
Bishkek 720024, Kyrgyz Republic}
\date{\today}

\begin{abstract}
The multidimensional gravity on the principal bundle with the 
{\rm SU}(2) gauge group is considered. The numerical investigation 
of the spherically symmetric metrics with the center of symmetry 
is made. The solution of the gravitational equations depends on the 
boundary conditions of the ``SU(2) gauge potential''  (off-diagonal metric 
components) at the symmetry center and on  the type of symmetry 
(symmetrical or antisymmetrical) of these potentials. 
In the chosen range of the boundary conditions it is shown that 
there are two types of solutions: wormhole-like and flux tube. 
The physical application of 
such kind of solutions as quantum handles in a spacetime foam 
is discussed. 
\end{abstract}

\maketitle
\section{Introduction} 

At the present time the most popular version of multidimensional (MD) 
gravity is a Kaluza-Klein gravity (for review see Ref. 
\cite{overduin:1997pn}) in which all space-like directions are 
equivalent, i.e. extra coordinates are the same as the space coordinates. 
Such approach is very natural but it 
has a great problem with the sharing\footnote{This means: a mechanism 
should exist in Nature which does some dimensions observed for us and remaining 
extra dimensions unobserved at least for the modern experimental physics} 
of extra dimensions (ED). 
In other words we should have some natural mechanism for the sharing  
the ED and space coordinates. It is well known that it is very difficult 
to realize such mechanism 
in the empty spacetime. It is necessary to introduce some external matter
field 
for such sharing of ED. Of course such way kills the 
Einstein's idea that the matter can be effectively constructed from the pure 
geometry.
\par 
There is another possibility for the renewal of  the 
above-mentioned Einstein's idea. We would like to return   to the initial   
interpretation of Kaluza-Klein theory in which that all physical quantities 
should not depend on the ED. How it can be done? We offer to take the ED 
as a gauge group (for example, U(1), SU(2) or SU(3) and so on). The 
advantages of such an  approach are obvious: (a) we will have the matter as
the 
gauge fields; (b) the ED make up a symmetric space and therefore the 
physical fields will not depend on the extra  coordinates. 
The first item (a) follows from the 
following theorem \cite{sal}, \cite{per}: 
\par
\textit{
Let $G$ be the group fibre of the principal bundle. Then 
there is a one-to-one correspondence between the $G$-invariant 
metrics $G_{AB}$ ($A,B$ are the multidimensional indices) 
on the  total  space ${\cal X}$
and the triples $(g_{\mu \nu }, A^{a}_{\mu }, \varphi)$.} 
\begin{equation}
ds^2 = G_{AB} dx^A dx^B = \varphi (x^\alpha) \sum^{\dim G}_{a=5}
\left [\sigma ^a + 
A^a_\mu (x^\alpha)dx^\mu \right ]^2 +
g_{\mu\nu}(x^\alpha) dx ^\mu dx^\nu
\label{intr-1}
\end{equation}
\textit{
here $g_{\mu \nu }$ is Einstein's pseudo  -
Riemannian metric on the base; $A^{a}_{\mu }$ is the gauge field 
of the group $G$ (the off-diagonal components of 
the multidimensional metric); 
$dl^2 = \varphi (x^\mu) \sigma^a\sigma_a$  is the 
symmetric metric on the fibre.} 
\par 
This theorem tells us that the inclusion the off-diagonal 
components of the MD metric is equivalent to the inclusion 
gauge fields (U(1), SU(2) or SU(3)) and 
a scalar field $\varphi(x^\mu)$ which is connected with the 
linear size of the extra dimensions. 
These geometrical fields can act as the source of the 
exotic matter 
necessary for the formation of the wormhole's (WH) mouth. Such 
solutions were obtained in Ref's.
\cite{chodos} \cite{clem} \cite{dzhsin} \cite{dzhhj}. These
solutions are spherically symmetric WH-like 
metrics\footnote{The difference between ordinary wormhole and 
wormhole-like metrics will be discussed in section \ref{disc}} 
with finite longitudinal size. The throat of these
WH-like solutions is located between two invariant surfaces 
on which $ds^2 = 0$. These results indicate that the exotic matter necessary 
for the formation of the WH mouth can 
appear in \textbf{\textit{vacuum multidimensional 
gravity}} from the off-diagonal elements of the metric
(the gauge fields) and from the metric on the fibre 
(the scalar field), rather than coming from some externally
given exotic matter. 
\par
The second item (b) leads to the fact that all physical fields 
(scalar field $\varphi$, gauge fields $A_\mu^a$ and 4D metrical tensor
$g_{\mu\nu}$) 
do not depend on the extra 
coordinates. Moreover the ED have an additional structure: every 
point is an element of a group (gauge group U(1), SU(2) or SU(3)), and 
\textit{\textbf{such structure cannot be destroyed 
by any perturbations.}} This means that not any physical particle can 
penetrate to the ED. Otherwise it will destroy the symmetry of the ED and 
consequently an algebraic  structure of the gauge group.
In fact the gravity on the principal bundle give us the natural 
way for the compactification and sharing of the ED. 

\section{Gravity Equations for SU(2) Gauge Group as the 
Extra Dimensions}

In this case the multidimensional (MD) 
spacetime is a total space of the principal bundle with 
SU(2) structural group. The fibre of this principal bundle (the ED) is SU(2) 
gauge group and the base is the 4D spacetime. The Langrangian is
\begin{equation}
  L = \sqrt{-G}R
  \label{sec2-10}
\end{equation}
where $G$ is the determinant and $R$ is the Ricci scalar 
of the MD metric. 
According to above-mentioned theorem the MD metric on the 
total space is
\begin{equation}
ds^2 = \Sigma _{\bar A} \Sigma ^{\bar A} 
\label{sec2-20}
\end{equation}
where 
\begin{eqnarray}
\Sigma ^{\bar A} & = & h^{\bar A}_B dx^B ,
\label{sec2-30} \\
\Sigma ^{\bar a} & = & \varphi(x^\alpha)
\left(\sigma ^{\bar a} + 
A^{\bar a}_\mu (x^\alpha)dx^\mu \right),
\label{sec2-40} \\
\sigma ^{\bar a} & = & h^{\bar a}_b dx^b , 
\label{sec2-50} \\
\Sigma ^{\bar \mu} & = & h^{\bar \mu}_\nu (x^\alpha)dx^\nu 
\label{sec2-60}
\end{eqnarray}
here $x^B$ are the coordinates on the total space; 
$B = 0,1,2,3,5,6,7$ is the MD index; $\dim SU(2) = 3$; 
$x^a$ is the coordinates on the group $SU(2)$ 
($a = 5,6,7$); 
$x^\mu = 0,1,2,3$ are the coordinates on the base of the bundle; 
$\alpha ,\mu ,\nu = 0,1,2,3$; 
$\bar A = (\bar \mu ,\bar a)$ is the sieben-bein index; 
$h^{\bar A}_B$ is the $N$-bein; 
$\sigma ^{\bar a} $ are the 1-forms on the  group $SU(2)$ satisfying  
$d \sigma ^{\bar a} = \epsilon^{\bar a}_{\bar b\bar c} 
\sigma ^{\bar b} \sigma ^{\bar c}$; 
($\epsilon^{\bar a}_{\bar b\bar c}$) are the structural 
constants for the gauge group $SU(2)$; the signature of the MD metric 
is $\eta_{\bar{A}\bar{B}} = (+,-,-,-,-,-,-)$. 
We must note that the functions $\varphi,A^{\bar a}_\mu, h^{\bar \mu}_\nu$
can depend only on the $x^\mu$ points on the base 
as  the fibres of  our bundle are homogeneous spaces. 
Varying with respect to our physical degrees of freedom \cite{dzhhj}
leads to the following equations system
\begin{eqnarray}
R^{\bar A}_\mu & = & 0 ,
\label{sec2-70} \\
R^{\bar a}_{\bar a} = 
R^{\bar 5}_{\bar 5} + R^{\bar 6}_{\bar 6} + R^{\bar 7}_{\bar 7}
& = & 0 .
\label{sec2-80} 
\end{eqnarray}
\par 
According to above-mentioned theorem the following dimensional reduction 
of the Ricci scalar $R(E)$ on the total space 
of the principal bundle (\cite{per}) 
\begin{eqnarray}
&&\int d^4 x d^d y \sqrt{ \vert \det G_{AB} \vert} R(E) = 
\nonumber \\
&&V_G \int d^4 x \sqrt{ \vert  g \vert } \varphi ^{d/2} 
\left [ R(M) + R(G) - \frac{1}{4} \varphi F^a_{\mu\nu}F_a^{\mu\nu} 
+ \frac{1}{4}d (d - 1) \partial _\mu \varphi \partial ^\mu \varphi 
\right ]
\label{sec2-90}
\end{eqnarray}
here $R(M), R(G)$ are the Ricci scalars of the base and structural group 
of the principal bundle respectively; $V_G$ is the volume of the group
$SU(2)$. 
The independent degrees of freedom in the gravity on the principal bundle 
are: scalar field $\varphi(x^\alpha)$, gauge potential $A^a_\mu(x^\alpha)$ 
and 4D metric tensor $g_{\mu\nu}(x^\alpha)$. The most important difference 
of this theory from the modern variants of the Kaluza - Klein gravity is that 
here we should vary on the $\varphi$ but not with the every component of the
metric on the ED ($G_{55}, G_{66}$ and so on). This leads to essential 
decreasing of the number of gravitational equations and to 
appearing of WH-like vacuum solutions which are necessary for the 
Einstein/Wheeler idea ``mass without mass'' and ``charge without charge''. 
For the ordinary 7D Kaluza-Klein theory with the Einstein equations 
$R_{AB} = \kappa T_{AB}$ we have $N = \frac{7\cdot (7+1)}{2} = 28$ 
equations. In our case the number of Eq's\eqref{sec2-70}, \eqref{sec2-80} 
are $N = (number\; of\; 4D\; Einstein\; eq's) + (number\; of\; Yang-Mills\; eq's)
 + (1\; eq.\; for\; scalar\; field) = 10 + 12 + 1 = 23$. Such essential 
simplification of the theory structure leads to the fact that in our case 
vacuum solutions (wormhole-like) appear which cannot exist in the ordinary 
Kaluza-Klein gravity (where we must destroy the primary Einstein idea about 
a geometrization of physics and insert a multidimensional matter). 
\par
Another remarkable peculiarity of such kind of gravitational theories is 
that the linear sizes of the ED can essentially differ from the 
Planck scale. It follows from the 
above-mentioned fact that not any physical test particle 
(and consequently not any physical body) can penetrate 
into the ED as far as it would destroy the algebraic  (consequently 
symmetric) structure of the ED. Therefore the essential property of the 
gravity on the principal bundle is that \textit{the linear size of the ED can be 
distinguished from the Planck scale.}
\par
And finally in these gravitational theory does not appear the 
compactification problem because the ED are the compact gauge group: U(1), 
SU(2), SU(3) or another compact Lie group. This means that in our approach 
the compactification problem is not dynamical one and the compactness of the 
ED is inserted initially in such kind of the MD theory.

\section{Metric Ansatz and Initial Equations}

We will search a solution for the following 7D metric 
(here we follow to Ref. \cite{dzhhj})
\begin{equation}
ds^2 = \frac{\Sigma ^2(r)}{u^3 (r)} dt^2 - dr^2 - a(r) 
\left(d\theta^2 + \sin ^2\theta d\phi ^2 \right) - 
R_0^2 u(r)\left (\sigma ^a + A^a_\mu dx^\mu \right )^2 
\label{sec3-10}
\end{equation}
here $r_0$ is some constant, $\sigma ^a$ ($a=5,6,7$) are the 
Maurer-Cartan form with relation 
$d\sigma ^a = \epsilon^a_{bc} \sigma ^b\sigma ^c$
\begin{eqnarray}
\sigma ^{1} & = & \frac{1}{2}
(\sin \alpha d\beta - \sin \beta \cos \alpha d\gamma ),
\label{sec3-20}\\
\sigma ^{2} & = & - \frac{1}{2}(\cos \alpha d\beta +
\sin \beta \sin \alpha d\gamma ),
\label{sec3-30}\\
\sigma ^{3} & = & \frac{1}{2}(d\alpha +\cos \beta d\gamma ),
\label{sec3-40}
\end{eqnarray}
where $0\le \beta \le \pi , 0\le \gamma \le 2\pi , 
0\le \alpha \le 4\pi $ are the Euler angles on the fibre. 
We choose the potential $A^a_\mu$ in the ordinary monopole-like form 
\begin{eqnarray}
A^{a}_{\theta } & = & \frac{1}{2}(1 - f(r))\{ \sin \phi ;-\cos \phi ;
0\} ,
\label{sec3-50}\\
A^{a}_{\phi } & = & \frac{1}{2}(1 - f(r))
\sin \theta \{\cos \phi \cos \theta ;
\sin \phi \cos \theta ;-\sin \theta \},
\label{sec3-60}\\
A^{a}_{t} & = & v(r)\{ \sin \theta \cos \phi ;
\sin \theta \sin \phi ;\cos \theta \},
\label{sec3-70}
\end{eqnarray}
Let us introduce the color electric $E^a_i$ and magnetic 
$H^a_i$ fields 
\begin{eqnarray}
E^a_i & = & F^a_{ti} ,
\label{sec3-80} \\
H^a_i & = & \sqrt \gamma \epsilon _{ijk} \sqrt {g_{tt}}F^{ajk}
\label{sec3-90}
\end{eqnarray}
here the field strength components are defined via 
$F^a_{\mu\nu} = A^a_{\nu ,\mu} - A^a_{\mu ,\nu} + 
\epsilon ^a_{bc}A^b_\mu A^c_\nu$, 
$\gamma$ is the determinant of the 3D space matrix, 
($i,j = 1,2,3$) are the space index. In our case we have
\begin{eqnarray}
E_r \propto v' , \qquad E_{\theta , \phi} \propto vf ,
\label{sec3-100} \\
H_r \propto \frac{\Sigma }{u^{3/2}}\frac{1 - f^2}{a} , 
\qquad H_{\theta , \phi} \propto f'
\label{sec3-110}
\end{eqnarray}
The substitution to the 7D gravitational equations 
\eqref{sec2-70} \eqref{sec2-80} leads to the following system 
of equations 
\begin{eqnarray}
\frac{\Sigma ''}{\Sigma} + \frac{a'\Sigma '}{a\Sigma} - 
\frac{4}{R_0^2u} - \frac{R_0^2u}{4a} {f'}^2 - 
\frac{R_0^2u}{8a^2}\left (f^2 -1 \right )^2 & = & 0 ,
\label{sec3-120} \\
24 \frac{\Sigma 'u'}{\Sigma u} - 24\frac{{u'}^2}{u^2} + 
16 \frac{a'\Sigma '}{a\Sigma} + 4 \frac{{a'}^2}{2a^2} - 
\frac{16}{a} + & & 
\nonumber \\
4 \frac{R_0^2u^4}{\Sigma^2}{v'}^2 - 
2\frac{R_0^2u}{a} {f'}^2 - 8 \frac{R_0^2u^4}{a\Sigma ^2}f^2v^2 + 
\frac{R_0^2u}{a^2}\left (f^2 -1 \right )^2 - 
\frac{48}{uR_0^2} & = & 0,
\label{sec3-130}\\
\frac{a''}{a} + \frac{a'\Sigma '}{a\Sigma} - \frac{2}{a} + 
\frac{R_0^2u}{4a} {f'}^2 - \frac{R_0^2u^4}{a\Sigma ^2}f^2v^2 + 
\frac{R_0^2u}{4a^2}\left (f^2 -1 \right )^2 & = & 0,
\label{sec3-140}\\
\frac{u''}{u} + \frac{u'\Sigma '}{u\Sigma} - 
\frac{{u'}^2}{u^2} + \frac{u'a'}{ua} - 
\frac{4}{R_0^2u} + & &  
\nonumber \\
\frac{R_0^2u^4}{3\Sigma^2}{v'}^2 - 
\frac{R_0^2u}{6a} {f'}^2 + 
\frac{2R_0^2u^4}{3a\Sigma ^2}f^2v^2 - 
\frac{R_0^2u}{12a^2}\left (f^2 -1 \right )^2 & = & 0,
\label{sec3-150} \\
v'' + v'\left (-\frac{\Sigma '}{\Sigma} + 4\frac{u'}{u} + 
\frac{a'}{a}\right ) - \frac{2}{a} vf^2 & = & 0 ,
\label{sec3-160}\\
f'' + f'\left (\frac{\Sigma '}{\Sigma} + 4\frac{u'}{u} \right ) 
+ 4\frac{u^3}{\Sigma ^2}fv^2 - \frac{f}{a} \left (f^2 -1 \right ) 
& = & 0 .
\label{sec3-170} 
\end{eqnarray}
The equations  \eqref{sec3-160} and \eqref{sec3-170} 
correspond to the ordinary ``Yang-Mills" equations after 
the dimensional reduction. 
\par 
In the consequence of the WH symmetry the functions 
$a(r), \Sigma (r)$ and $u(r)$ are symmetric and 
the functions 
$f(r)$ and $v(r)$ can be either symmetric or antisymmetric. 
Consequently we would like to consider the following different cases
\begin{enumerate}
  \item 
  the function $f(r)$ is symmetric and the function $v(r)$ is 
  antisymmetric;
  \item 
  the function $f(r)$ is antisymmetric and the function $v(r)$ is 
  symmetric;
  \item 
  both functions $f(r)$ and $v(r)$ is antisymmetric; 
  \item
  both functions $f(r)$ and $v(r)$ is symmetric.
\end{enumerate}
The first case was considered in the Ref. \cite{dzhhj}. The result is 
that there are three type of solutions : wormhole-like, infinite and finite 
flux tubes solutions. 

\section{Boundary Conditions for the 4D Metric}

Now we will write the initial conditions for the 4D metric functions. 
At the center of symmetry of the WH ($r=0$) we can expand functions 
$a(r), \Sigma(r)$ and $u(r)$ by this manner 
\begin{eqnarray}
a(x) & = & 1 + \frac{a_2}{2}x^2 + \cdots ,
\label{sec4-10} \\
\Sigma (x) & = & \Sigma _0 + \frac{\Sigma _2}{2}x^2 + 
\cdots ,
\label{sec4-20}\\
u(x) & = & u_0 + \frac{u_2}{2}x^2 + \cdots ,
\label{sec4-30}
\end{eqnarray}
here we introduce the dimensionless coordinate $x = r/\sqrt{a_0}$ 
($a_0 = a(0)$) and redefine $a(x)/a_0 \to a(x)$, 
$\sqrt{a_0} v(x) \to v(x)$, $R_0^2/a_0 \to R_0^2$. Then we can rescale 
time and the constant $R_0$ so that $\Sigma _0 = u_0 = 1$. Thus we 
have the following initial conditions for the functions 
$a(r), \Sigma(r)$ and $u(r)$ for the numerical calculations 
\begin{eqnarray}
a_0 = 1, \qquad u_0 = 1, \qquad \Sigma _0 = 1, 
\label{sec4-40} \\
a'_0 = 0, \qquad u'_0 = 0, \qquad \Sigma '_0 = 0. 
\label{sec4-50}
\end{eqnarray}

\section{Function $f(r)$ is antisymmetric and function $v(r)$ is 
symmetric}

\begin{figure}
\begin{center}
\fbox{
\includegraphics[height=7cm,width=7cm]{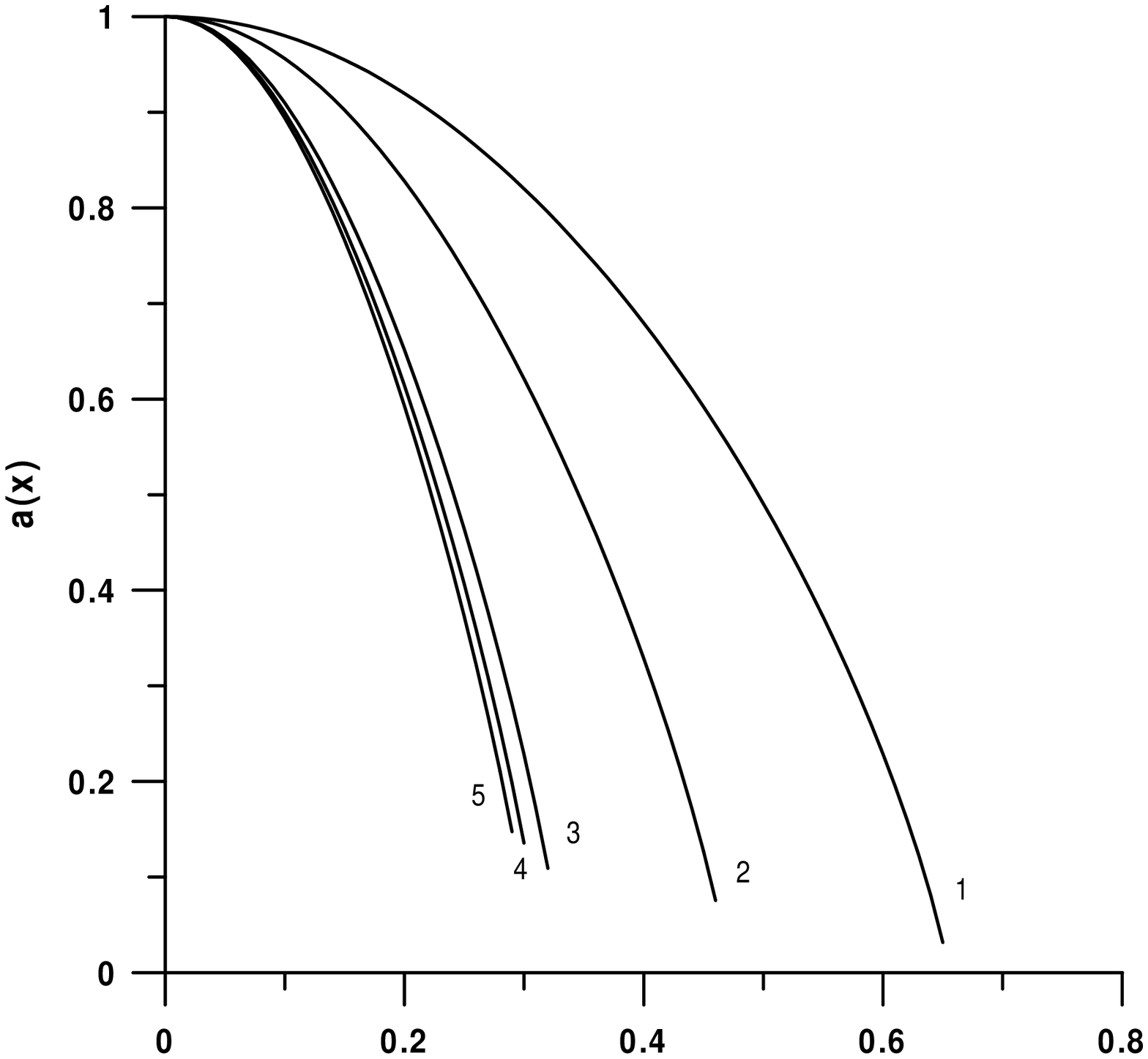}}
\caption{The case 2 : the function $a(x)$. 
$v_0 = 0.2$;
$f_1 = 0.3,\; 0.5,\; 0.6,\; 0.61,\; 0.615$}
\label{cs2-ax}
\end{center}
\end{figure}

\begin{figure}
\begin{center}
\fbox{
\includegraphics[height=7cm,width=7cm]{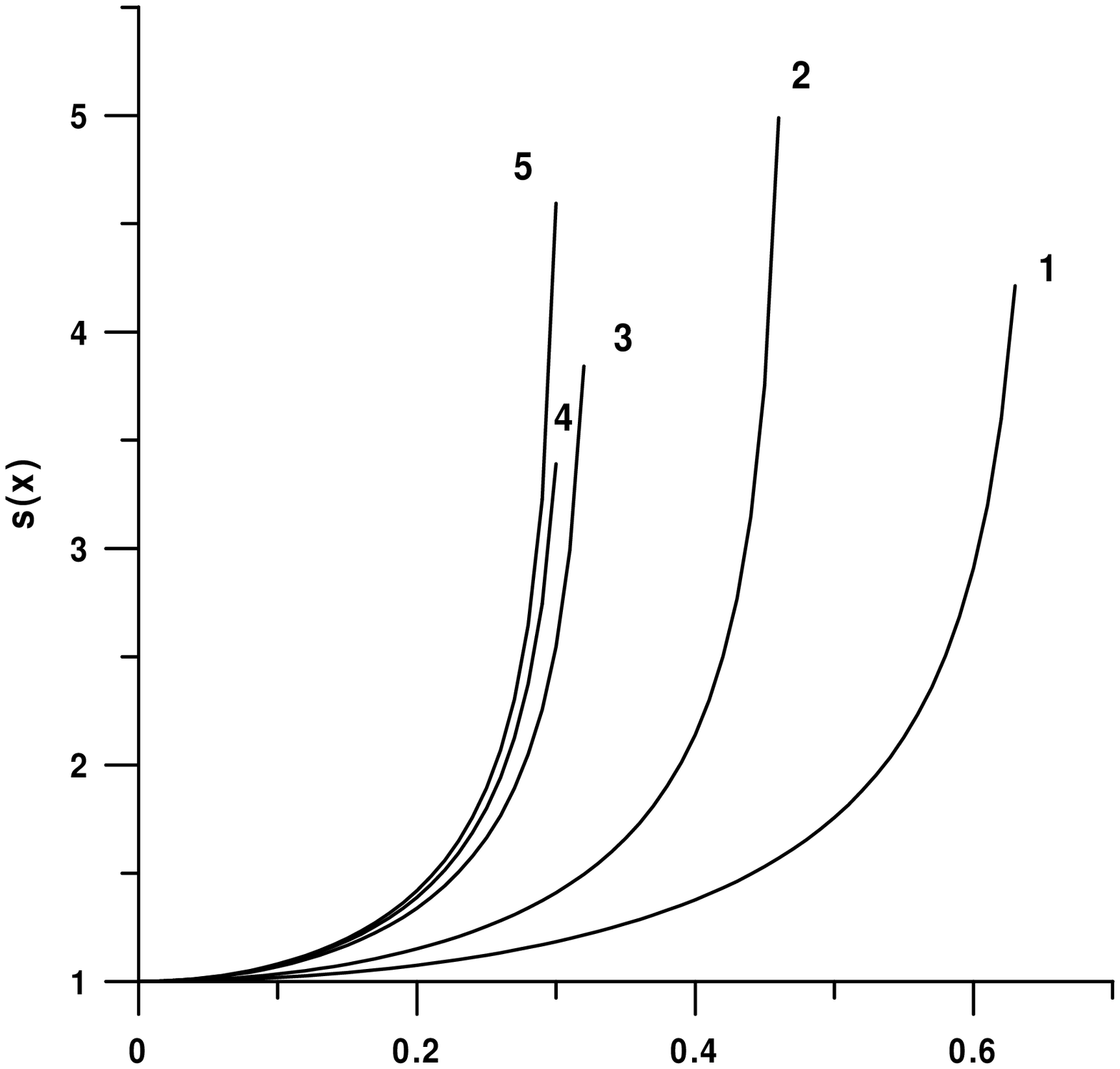}}
\caption{The case 2 : the function $s(x)$.
$v_0 = 0.2$;
$f_1 = 0.3,\; 0.5,\; 0.6,\; 0.61,\; 0.615$}
\label{cs2-sx}
\end{center}
\end{figure}

\begin{figure}
\begin{center}
\fbox{
\includegraphics[height=7cm,width=7cm]{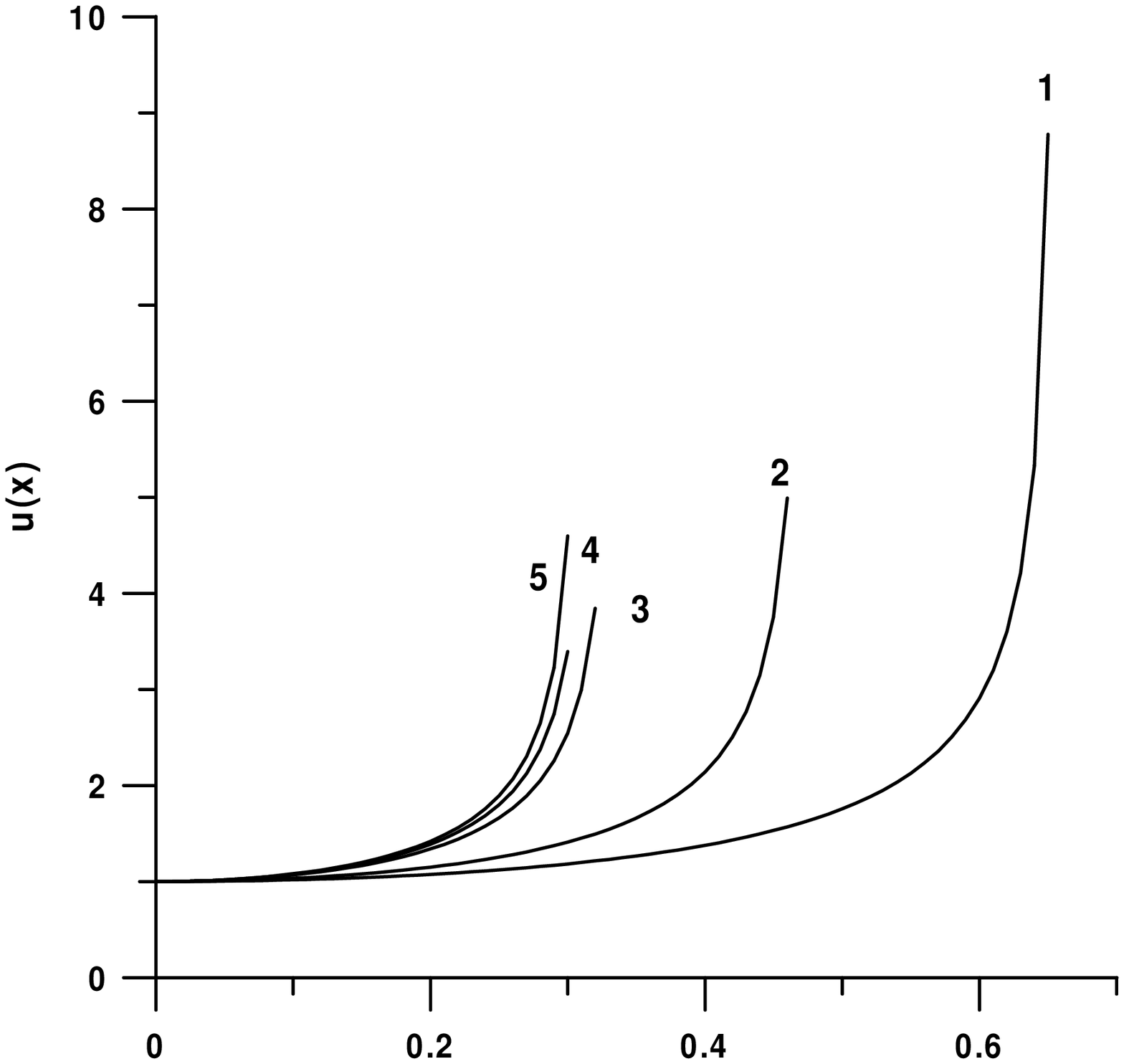}}
\caption{The case 2 : the function $u(x)$.
$v_0 = 0.2$;
$f_1 = 0.3,\; 0.5,\; 0.6,\; 0.61,\; 0.615$}
\label{cs2-ux}
\end{center}
\end{figure}

\begin{figure}
\begin{center}
\fbox{
\includegraphics[height=7cm,width=7cm]{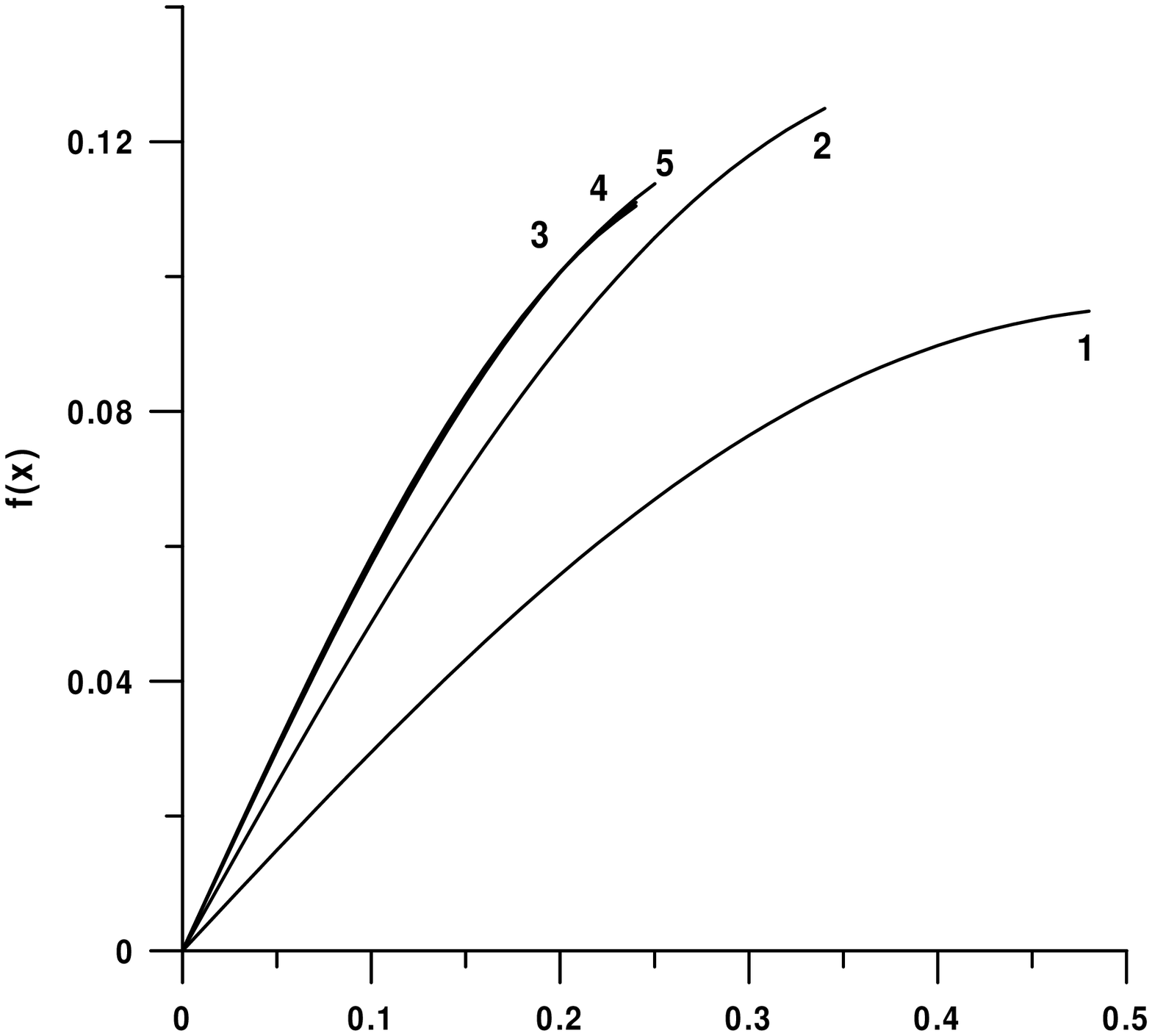}}
\caption{The case 2 : the function $f(x)$.
$v_0 = 0.2$;
$f_1 = 0.3,\; 0.5,\; 0.6,\; 0.61,\; 0.615$}
\label{cs2-fx}
\end{center}
\end{figure}

\begin{figure}
\begin{center}
\fbox{
\includegraphics[height=7cm,width=7cm]{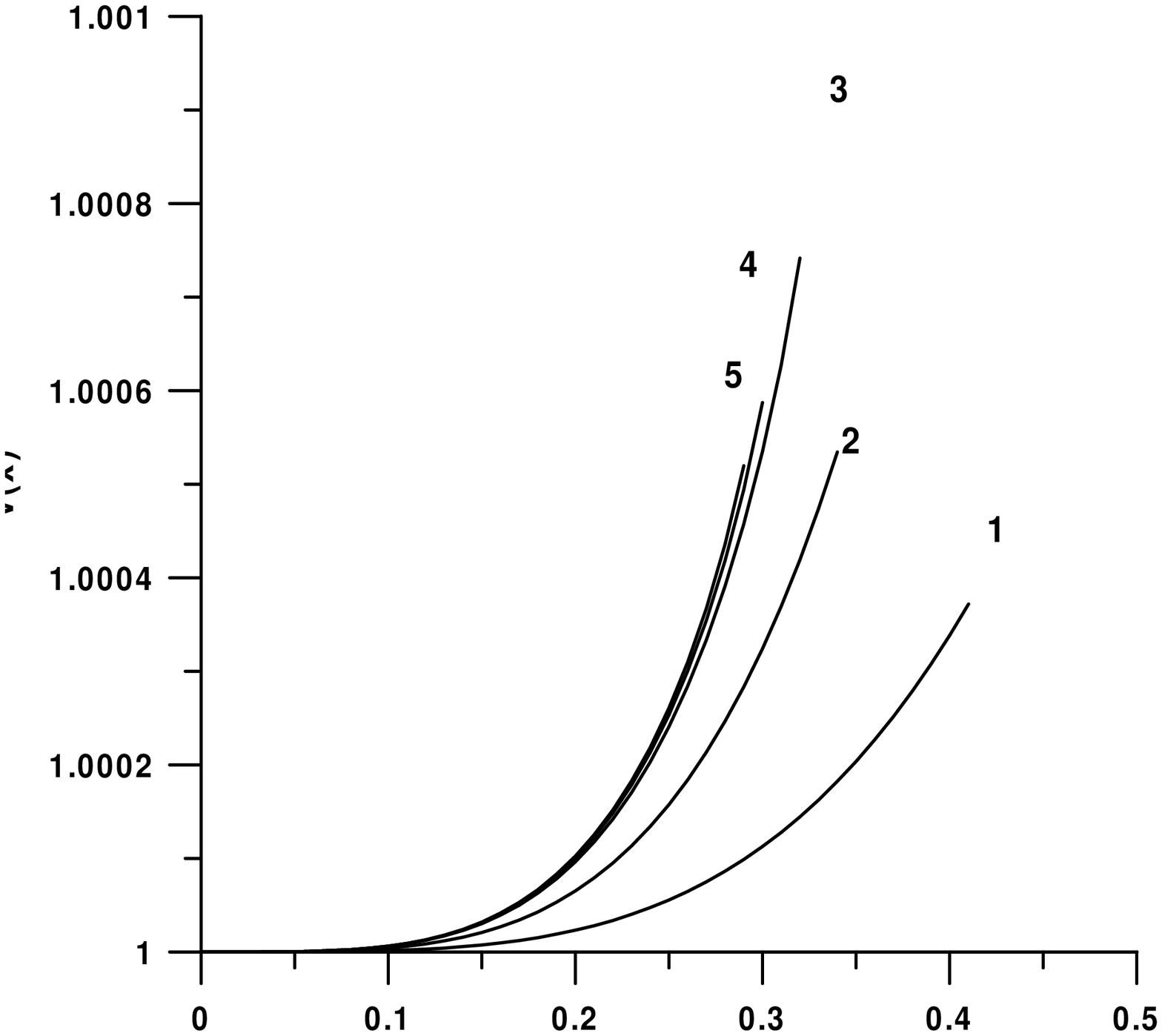}}
\caption{The case 2 : the function $v(x)$.
$v_0 = 0.2$;
$f_1 = 0.3,\; 0.5,\; 0.6,\; 0.61,\; 0.615$}
\label{cs2-vx}
\end{center}
\end{figure}

In this case we have the following expansion of functions 
$f(x)$ and $v(x)$ at the origin
\begin{eqnarray}
v(x) & = & v_0 + \frac{v_2}{2} x^2 +\cdots ,
\label{sec5-10}\\
f(x) & = & f_1 x + \frac{f_3}{6}x^3 + \cdots ,
\label{sec5-20}
\end{eqnarray}
this equation is written  in  dimensionless variables. 
The constrained equation \eqref{sec3-130} for the initial data 
give us 
\begin{equation}
R_0^2 = 4 \frac{2 + \sqrt{7 - 6 f_1^2}}{1 - 2f_1^2}
\label{sec5-30}
\end{equation}
As $R_0^2 > 0$ we have the following constraint for $f_1$ : 
$|f_1| < 1/\sqrt{2}$. The numerical calculations are presented 
on the Fig's \eqref{cs2-ax}, \eqref{cs2-sx}, \eqref{cs2-ux}, 
\eqref{cs2-fx}, \eqref{cs2-vx}. 
\par
\begin{figure}
\begin{center}
\fbox{
\includegraphics[height=7cm,width=7cm]{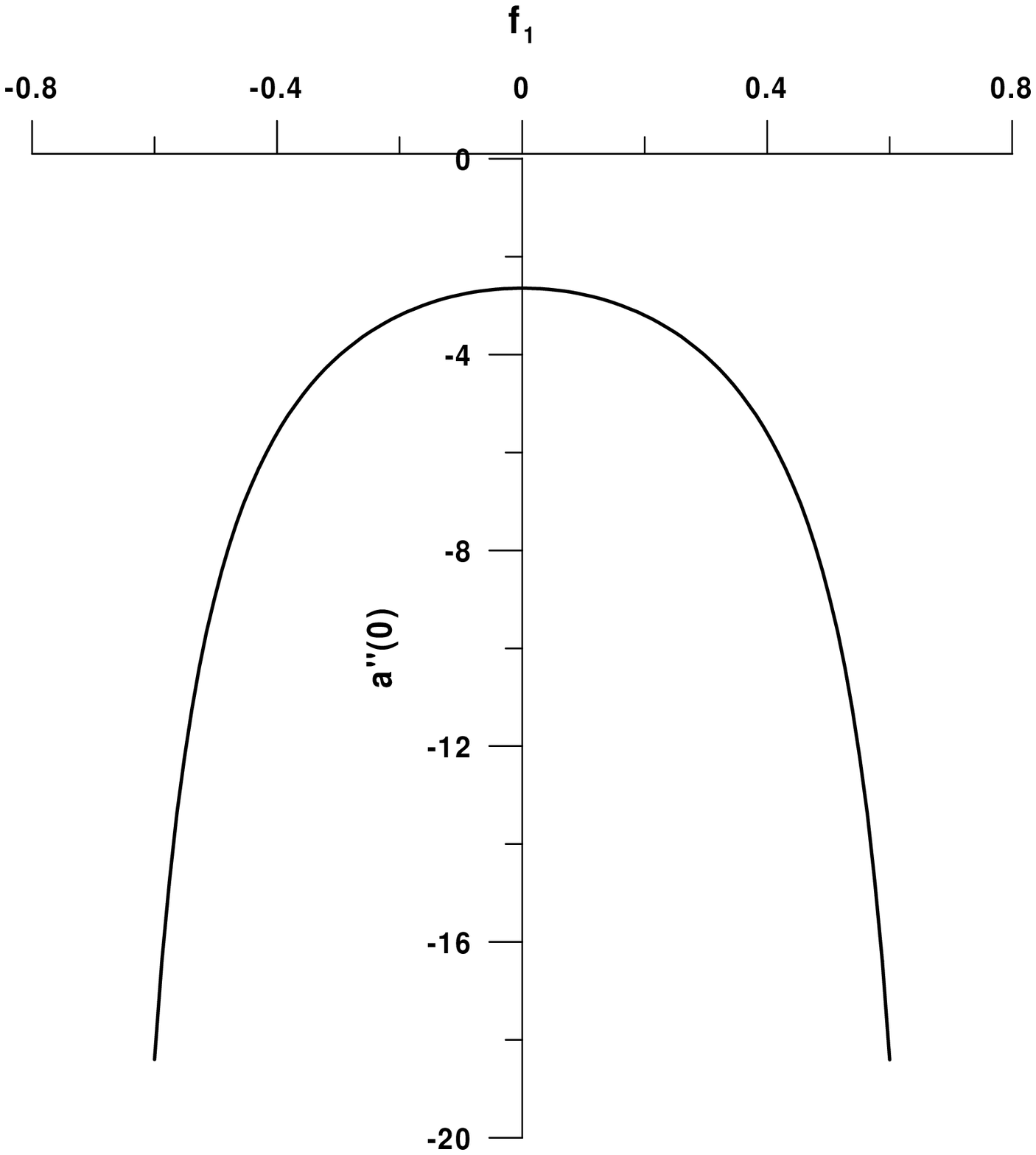}
}
\caption{The case 2 : the function $f(r)$ is antisymmetric and $v(r)$ 
is symmetric. $a_0''$ as a function on $f_1$.}
\label{cs2-rstr}
\end{center}
\end{figure}

 
In this case the numerical calculations show us that the 
$a(x)$ function is monotonically decreasing one and consequently 
its behaviour is defined with the value of 
$a''_0(0)$. Eq. \eqref{sec3-140} give us the following expression 
\begin{equation}
\frac{a''_0}{a_0} = 2 - \left(1 + f_1^2\right) 
\frac{2 + \sqrt{7 - 6 f_1^2}}{1 - 2f_1^2}
\label{sec5-40}
\end{equation}
On the Fig.\eqref{cs2-rstr} this curve is shown. As a result we see that 
$a''_0 < 0$ for all $f_1$ and consequently it confirms our 
numerical investigation that $a(x)$ is monotonically decreasing 
function. 

\section{Both Functions $f(r)$ and $v(r)$ are antisymmetric}

In this case we have the following expansion of functions 
$f(x)$ and $v(x)$ at the origin
\begin{eqnarray}
v(x) & = & v_1 x + \frac{v_3}{6} x^3 +\cdots ,
\label{sec6-10}\\
f(x) & = & f_1 x + \frac{f_3}{6}x^3 + \cdots ,
\label{sec6-20}
\end{eqnarray}
this equation is written  in  dimensionless variables. 
The constrained equation (\ref{sec3-130}) for the initial data 
give us 
\begin{equation}
r_0^2 = 4 \frac{2 + \sqrt{7 + 12 v_1^2 - 6 f_1^2}}
{1 + 4 v_1^2 - 2 f_1^2}.
\label{sec6-30}
\end{equation}
The positivity  condition of $R_0^2$ give us 
\begin{equation}
  2 f_1^2 - 4 v_1^2 < 1 
\label{sec6-50}  
\end{equation}
\par
\begin{figure}
\begin{center}
\fbox{
\includegraphics[height=7cm,width=5cm]{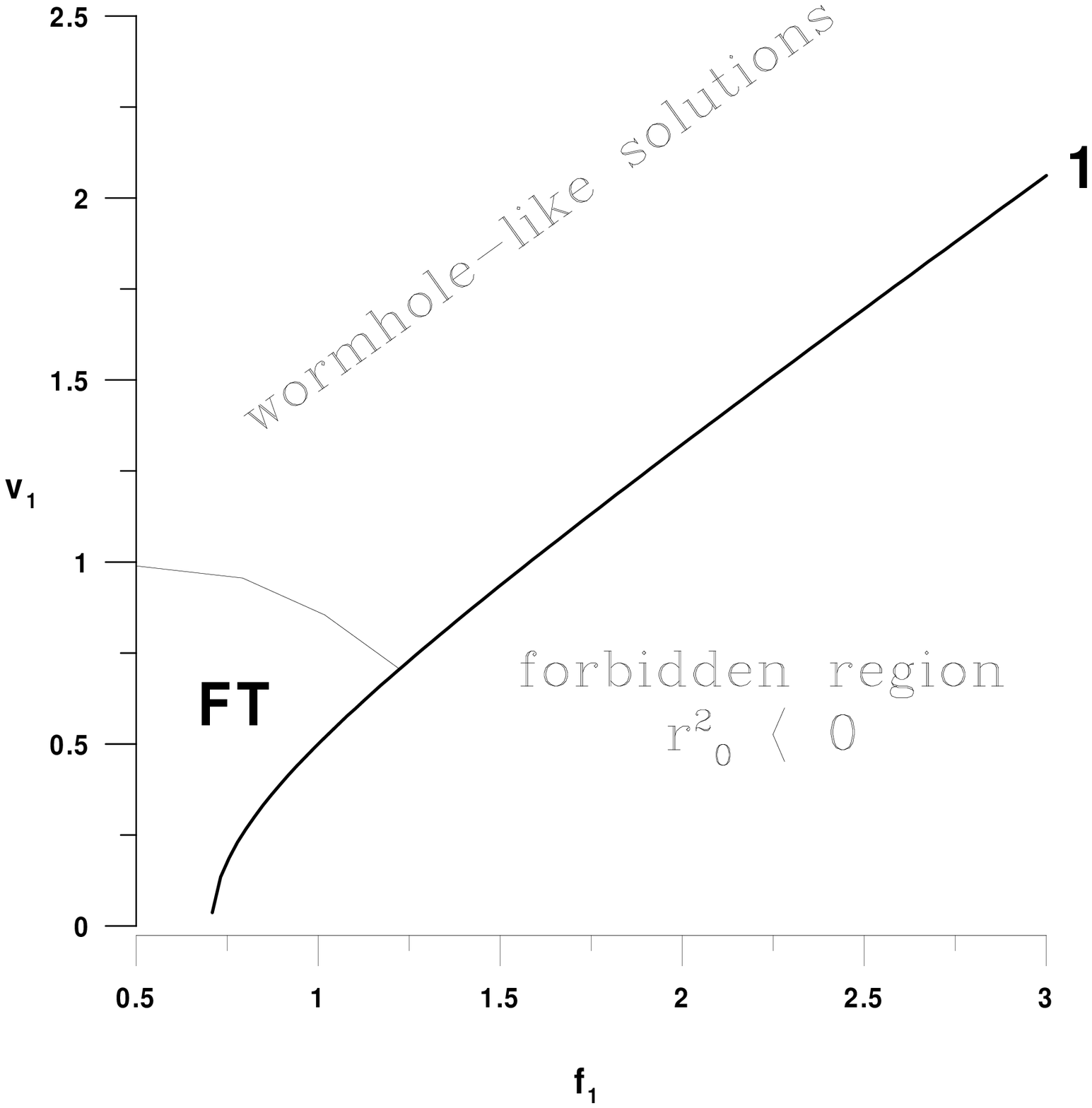}}
\caption{The case 3 : both functions $f(r)$ and $v(r)$ 
are antisymmetric. 
The \textbf{FT} region is the region with the flux tube  
solutions. $r_0^2 < 0$ for the region under curve 1}
\label{cs3-rstr}
\end{center}
\end{figure}

 
In the chosen range of the parameters $(f_1,v_1)$ 
the numerical calculations lead to the fact that on the 
$(f_1, v_1)$ plane there are two regions with 
the different solutions type. On Fig.\eqref{cs3-rstr} these 
regions with the  different type of solutions are shown. 
\par
\begin{figure}
\begin{center}
\fbox{
\includegraphics[height=7cm,width=7cm]{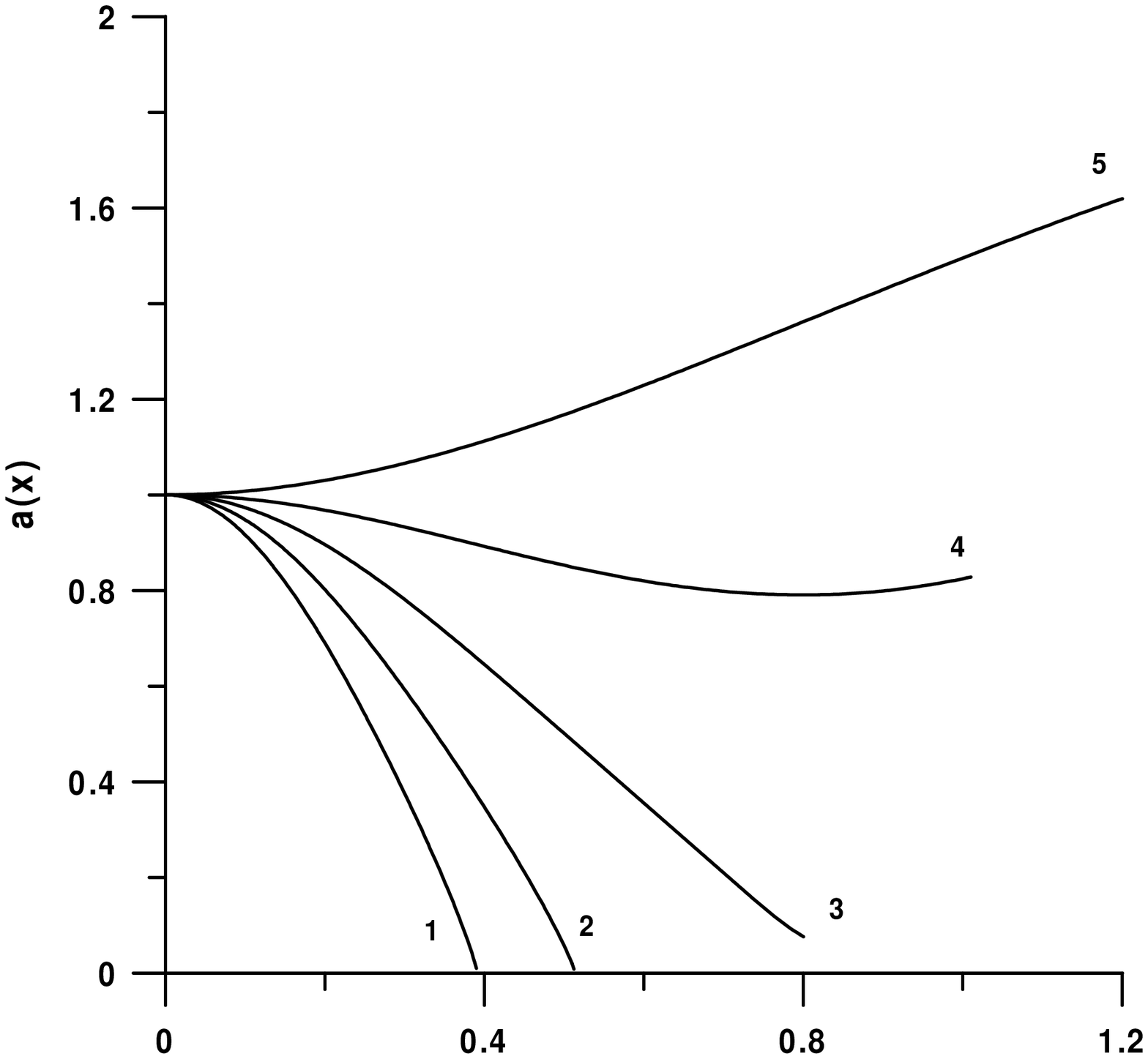}}
\caption{The case 3 : the function $a(x)$. 
$f_1 = 1.0$; $v_1 = 0.6,\; 0.65,\; 0.75,\; 1.0,\; 5.0$ 
according to the curves 1, 2, 3, 4, 5.}
\label{cs3-ax}
\end{center}
\end{figure}

\begin{figure}
\begin{center}
\fbox{
\includegraphics[height=7cm,width=7cm]{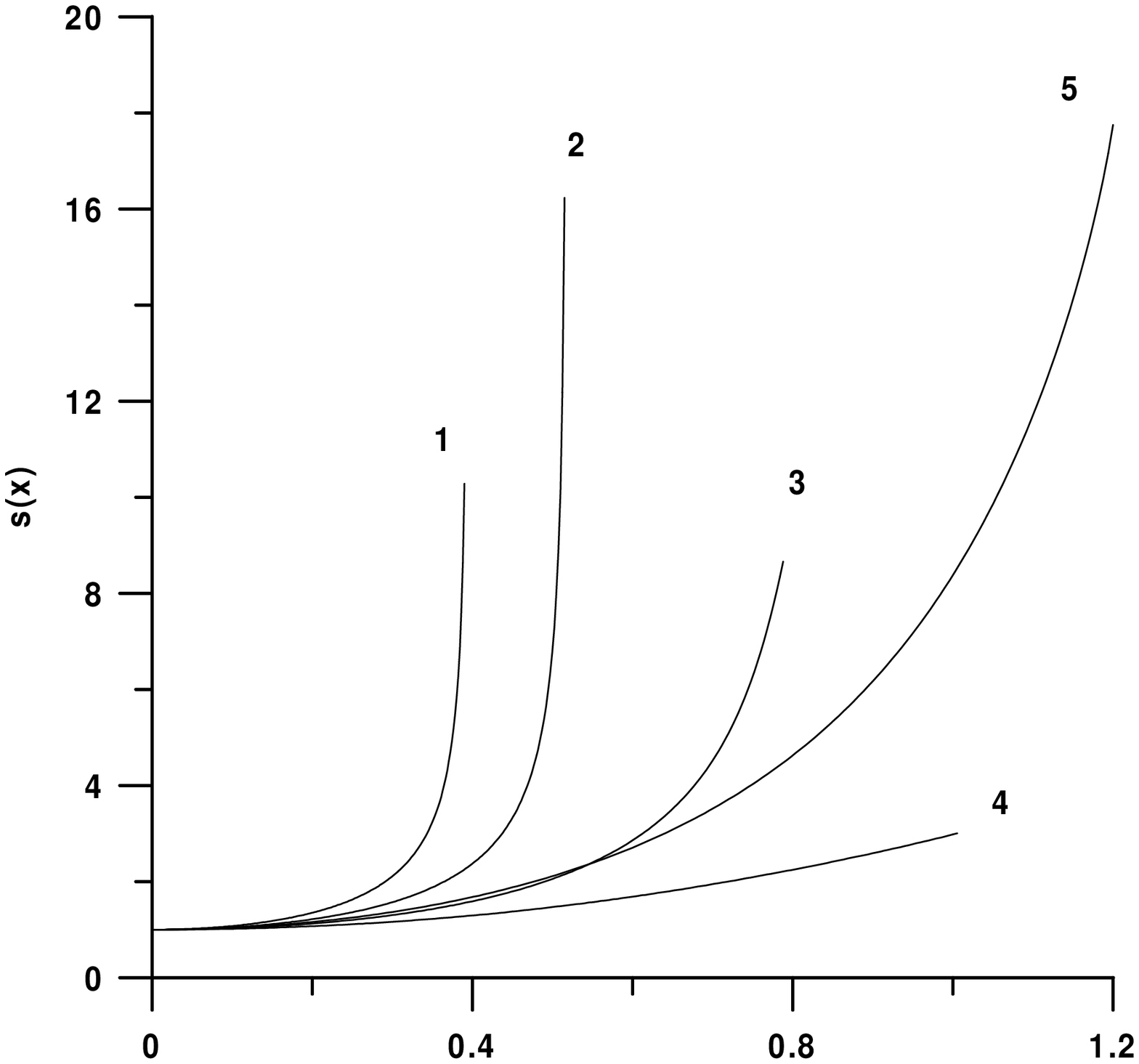}}
\caption{The case 3 : the function $s(x)$.
$f_1 = 1.0$; $v_1 = 0.6,\; 0.65,\; 0.75,\; 1.0,\; 5.0$}
\label{cs3-sx}
\end{center}
\end{figure}

\begin{figure}
\begin{center}
\fbox{
\includegraphics[height=7cm,width=7cm]{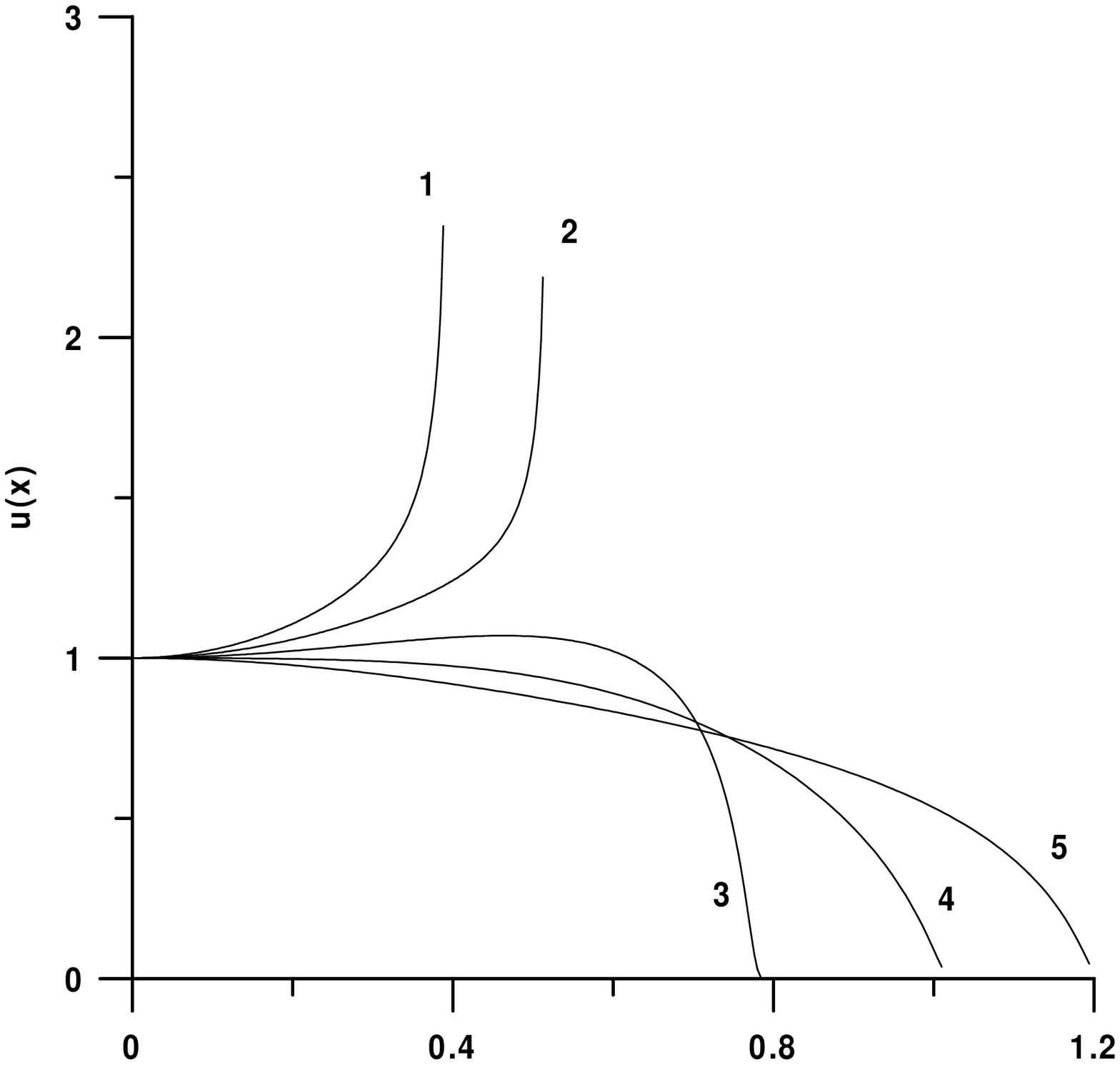}}
\caption{The case 3 : the function $u(x)$.
$f_1 = 1.0$; $v_1 = 0.6,\; 0.65,\; 0.75,\; 1.0,\; 5.0$}
\label{cs3-ux}
\end{center}
\end{figure}

\begin{figure}
\begin{center}
\fbox{
\includegraphics[height=7cm,width=7cm]{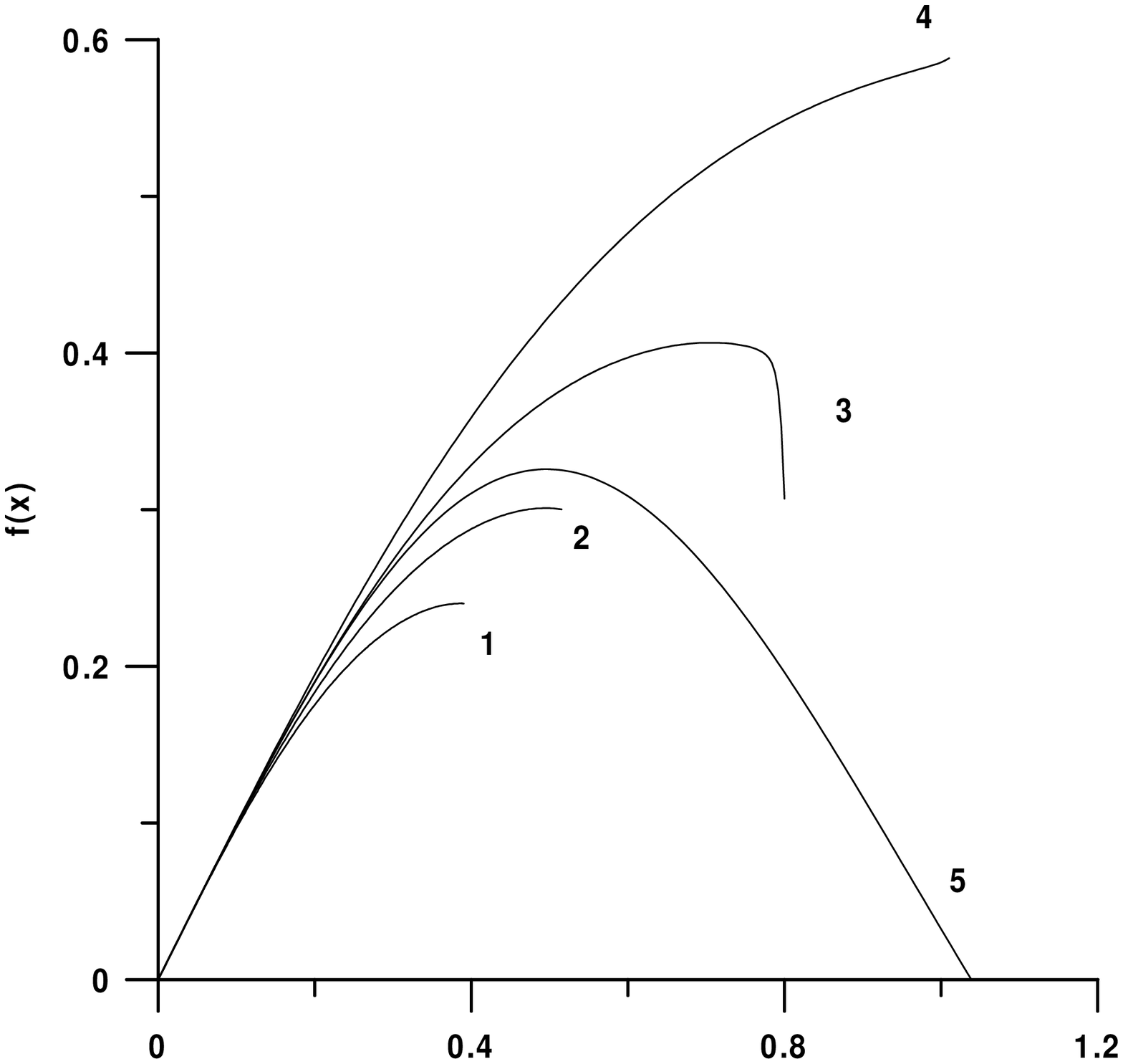}}
\caption{The case 3 : the function $f(x)$.
$f_1 = 1.0$; $v_1 = 0.6,\; 0.65,\; 0.75,\; 1.0,\; 5.0$}
\label{cs3-fx}
\end{center}
\end{figure}

\begin{figure}
\begin{center}
\fbox{
\includegraphics[height=7cm,width=7cm]{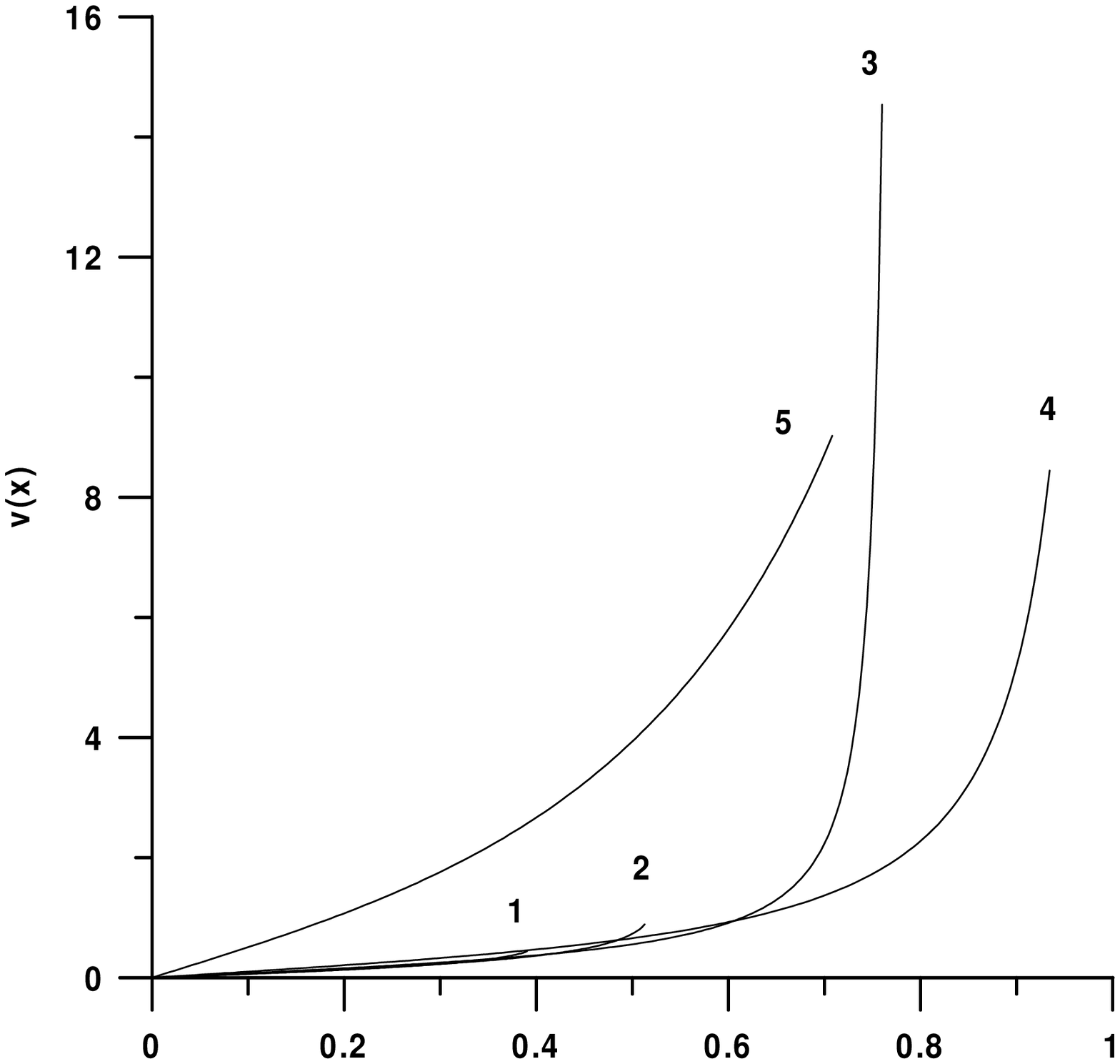}}
\caption{The case 3 : the function $v(x)$.
$f_1 = 1.0$; $v_1 = 0.6,\; 0.65,\; 0.75,\; 1.0,\; 5.0$}
\label{cs3-vx}
\end{center}
\end{figure}

 
The numerical calculations are presented on the Fig's 
\eqref{cs3-ax}, \eqref{cs3-sx}, \eqref{cs3-ux}, 
\eqref{cs3-fx}, \eqref{cs3-vx}. We see that there is two 
type of solutions. The first type with decreasing $u(x)$ we can 
name as a WH-like solutions. For this type of solutions we have 
the following condition $a_0'' > 0$. 
With an accuracy of the numerical 
calculations we can say that there is a point $x_1$ for which 
$u(\pm x_1) = 0$. Probably in these points $ds^2 = 0$ that is 
similar to the 5D case investigated in Ref.\cite{dzhsin}. 
We intend to investigate in more details the behavior of the 
metric close to such points with the help of approximate analytical 
methods in the future. The second type satisfies the condition 
$a_0'' < 0$ and with an accuracy of the numerical calculations 
there is a point $x_1$ for which $a(\pm x_1) = 0$. With 
great probability we have singularities in these two points, 
\textit{i.e.} such solutions are like to flux tube of color 
``electric'' and ``magnetic'' fields between two singularities. 
It is interesting to note that this type of solutions is very similar 
to the confinement mechanism in QCD where two quarks are located 
at the ends of a flux tube with color electric and magnetic fields 
running quark and antiquark. 

\section{Both Functions $f(r)$ and $v(r)$ are symmetric}

In this case we have the following expansion of functions 
$f(x)$ and $v(x)$ at the origin
\begin{eqnarray}
v(x) & = & v_0 + \frac{v_2}{2} x^2 +\cdots ,
\label{sec7-10}\\
f(x) & = & f_0 + \frac{f_2}{2}x^2 + \cdots ,
\label{sec7-20}
\end{eqnarray}
this equation is written  in  dimensionless variables. 
The constrained equation \eqref{sec3-130} for the initial data 
give us 
\begin{equation}
R_0^2 = 4 \frac{2 + \sqrt{4 + 3
\left[
\left(f_0^2 - 1\right)^2 - 8 f_0^2 v_0^2
\right]}}{\left(f_0^2 - 1\right)^2 - 8 f_0^2 v_0^2}
\label{sec7-30}
\end{equation}
The positivity  condition of $R_0^2$ gives us 
\begin{equation}
  \left(f_0^2 - 1\right)^2 > 8f_0^2 v_0^2
\label{sec7-40}  
\end{equation}
\par 
In this case the numerical calculations show us that the 
$a(x)$ function is monotonically decreasing one and consequently 
its behavior is defined with the value of 
$a''_0(0)$. Eq's \eqref{sec3-130} \eqref{sec3-140} give us the 
following expression 
\begin{equation}
  a''_0 = -\frac{r_0^2}{8}\left(f_0^2 - 1\right)^2 - 
  \frac{6}{r_0^2} < 0.
\label{sec7-50}
\end{equation}
As a result we see that 
$a''_0 < 0$ for all $f_0, v_0$ and consequently it confirms our 
numerical investigation that $a(x)$ is monotonically decreasing 
function. The qualitative behavior of all functions 
$a(x), \Sigma(x), u(x), f(x)$ and $v(x)$ are similar with the 
second case (where $f(x)$ is antisymmetric and $v(x)$ is  
symmetric one). 

\section{Discussion}
\label{disc}

On the basis of Ref.\cite{dzhsin} and the investigations presented 
here we can say that the spherically symmetric solutions of the gravity 
on the principal bundle with SU(2) gauge group as the structural 
group (extra dimensions) and which are symmetric regarding to some 
hypersurface $r=const$ (in our case $r=0$) are presented by three 
types :
\begin{enumerate}
  \item 
  Wormhole - like solutions. These solutions are characterized that 
  $a(0) = minimum$ and there are two hypersurfaces by $r = \pm r_0$ 
  where $ds^2 = 0$.
  \item 
  Infinite flux tube. This solution was presented in Ref.\cite{dzhsin2}. 
  It has $a(r) = const$ ($r \in (-\infty , +\infty)$) and color electric 
  and magnetic fields are parallel.
  \item 
  Finite flux tubes. These solutions have two singularities at 
  $r = \pm r_0$ and function $a(r)$ has a maximum at $r = 0$.
\end{enumerate}
The difference between these possibilities are determinated by 
the symmetry of $f(r)$ and $v(r)$ functions and their boundary 
conditions $f(0), f'(0), v(0)$ and $v'(0)$. It means that the type 
of solutions is completely defined by color electric and magnetic 
fields at the symmetry center of the metric.
\par
Now we would like to compare our designations for the solutions 
(wormhole and flux tube) with other similar definitions. 

\subsection{Wormholes}

According to Ref.\cite{visser} the wormhole is time independent 
(in the simplest case), nonrotating, and spherically symmetric 
bridge between two universes. The 4D spacetime metric can be put 
into the form 
\begin{equation}
  ds^2 = \frac{dt^2}{\Delta(r)} - dr^2 - a(r) 
  \left(
  d\theta ^2 + \sin ^2 \theta d\varphi^2
  \right)
\label{secwh-1}
\end{equation}
that can be compared with 4D part of Eq.\eqref{sec3-10}. The WH should 
have the following properties : 
\begin{itemize}
  \item 
  The coordinate $r \in (-\infty , +\infty)$. 
  \item 
  The event horizon is absent.
  \item 
  There is a throat where $a(0) = minimum$.
  \item
  There are two asymptotically flat regions at $r = \pm \infty$.
\end{itemize}
Comparing these definitions with our calculations we can see why 
we called our first type of solutions as \textit{wormhole-like} 
solutions : 
\begin{itemize}
  \item 
  The coordinate $r \in (-r_0 , +r_0)$. It is possible that 
  similar to the 5D wormhole-like solutions we can continue these 
  solutions up to $r \in (-\infty , +\infty)$. In this case the 
  metric signature will be changed : from $(+,-,-,-,-)$ by 
  $r \in (-r_0 , +r_0)$ to $(-,-,-,-,+)$ by $r \in (-\infty , -r_0)$
  and $r \in (+r_0 , +\infty)$. Such situation for 5D case was 
  discussed in Ref.\cite{dzhhj2}. Following to Bronnikov \cite{bron} 
  we can name   two hypersurfaces $r = \pm r_0$ where $ds^2 = 0$ and such 
  exchange of the metric signature takes place as $T-$horizons. 
  \item 
  There are two $T-$horizons.
  \item 
  There is a minimum of $a(r)$.
  \item
  The existence of asymptotically flat regions is not yet unknown but the 
  comparison with the 5D case give us a hope that such regions 
  exist.
\end{itemize}
Thus the distinction and similarities between these standard and our 
metrics is evident. The most important peculiarity of presented here 
wormhole-like solutions is that they are vacuum solutions without any 
matter. The color electric and magnetic SU(2) gauge fields are the 
off-diagonal components of 7D metric. 

\subsection{Euclidean wormholes}

Euclidean wormholes have the topology $R \times S^{n-1}$, 
where $R$ is an imaginary time and $S^{n-1}$ is $(n-1)$ dimensional 
sphere. They are commonly thought of as ``instantons'' in the 
gravitational field. The distinction with our case is evident.

\section{Flux tubes}

The notion of flux tube has arisen in the quantum chromodynamics where 
this one is a hypothesized tube filled with gauge field and stretched 
between quark-antiquark. Such field configuration arises in the 
consequence of a specific non-linear potential in the SU(3) 
Yang-Mills theory. 
\par 
In the gravity can be  some types of spacetime with a finite flux 
of electric/magnetic fields. 
\begin{itemize}
  \item 
  Melvin type of solutions (see, for example, Melvin Universe 
  \cite{melvin}). The $t=const$ spacelike section has the cylindrical 
  symmetry and filled with cylindrically symmetric magnetic field 
  which has a finite flux of this field. 
  \item 
  The modern multidimensional Kaluza-Klein theories have solutions which 
  can contain branes with some generalization of electric field and finite 
  flux of this fileds, \textit{i.e.} flux branes.
  \item
  Levi-Civita-Betotti-Robinson solutions \cite{Levi-Civita}. 
  These solutions can be multidimensional \cite{dzhsin2}. 4D part of these 
  metrics has the following topologies : 
  $R \times I \times S^2 = (time) \times (radial\; coordinate) \times (sphere)$. 
  The range of the radial coordinate can be finite $I = [-r_0 , +r_0]$ and 
  infinite $I = R = (-\infty , +\infty)$. Roughly speaking these solutions 
  have a finite cross section $(S^2)$. They are filled with 
  electric/magnetic field with finite flux of these fields. We have shown that 
  such solutions exist in our case and the difference between wormhole-like 
  and flux tube solutions is defined by the relation between electric and 
  magnetic fields at the center of symmetry of this metric. 
\end{itemize}

\section{Conclusions}

In this paper we have shown (with an accuracy of the numerical 
calculations) that the spherically symmetric metric of the 
MD gravity on the principal bundle with SU(2) gauge group as 
the ED are either WH-like or flux tube metrics. The numerical 
calculations indicate that the type of solutions is determined 
by a relation between the color ``electric'' ($E_{r,\theta,\varphi}$), 
``magnetic'' ($H_{r,\theta,\varphi,}$) fields, 
the scalar curvatures of the ED ($R_{ED}$) and the sphere $S^2$ 
($R_{S^2}$) at $r = 0$. 
Let us write the constraint equation \eqref{sec3-130} 
\begin{equation}
\begin{split}
    \frac{u_0^3}{\Sigma_0^2}{v'_0}^2 + \frac{1}{4a_0^2}
    \left(f_0^2 - 1\right)^2 - 
    \frac{1}{2a_0}{f'_0}^2 - 2\frac{u_0^3}{a_0\Sigma_0^2}f_0^2v_0^2 - 
    \frac{2}{r_0^2u_0}
    \left(\frac{2}{a_0} + \frac{6}{u_0r_0^2} \right) & = \\
    \left(E^2_r \right)_0 + 
    \frac{1}{4}\left(H_r^2\right)_0 - 
    2\left(E_\theta ^2\right)_0 - 
    \frac{1}{2}\left(H_\theta ^2\right)_0 - 
    \frac{2}{r_0u_0}\left(R_{S^2} + R_{ED}\right) & = 0 .
\label{ccl-1}
\end{split}
\end{equation}
Immediately we see that the necessary condition for the 
existence of the WH-like solutions is that it must be a flux of 
color ``electric'' $\Phi_E = 4\pi a_0 E_r^2$ 
and ``magnetic'' fields $\Phi_H = 4\pi a_0 H_r^2$ through the WH mouth. 
Only in this case Eg. \eqref{sec3-130} is fulfilled. The numerical
calculations 
confirm such conclusion: only for the first and the third cases 
with $E_r \propto f_0' = f_1 \neq 0$ we will have the WH-like solutions. 
\par 
The WH-like solutions can have  very interesting physical consequences. 
Let us imagine that the universe is separated into the regions with 
dynamical and nondynamical metric on the ED \cite{dzhhj}. 
It means that in the region 
with nondynamical $G_{\bar{a}\bar{a}}$ ($\bar{a} = 5,6,7$) metric 
components we have 4D gravity interacting with the SU(2) gauge field 
and we have not an equation for the scalar field $\varphi$ 
($\varphi(x^\mu) = const$) (this is the Kaluza - Klein gravity in its 
initial interpretation). In the region with dynamical 
$G_{\bar{a}\bar{a}}$ we have the full MD gravity. Such situation can be 
realized on the level of the spacetime foam, \textit{i.e.} the quantum 
handles of spacetime foam are the regions with the dynamical 
$G_{\bar{a}\bar{a}}$ and the region between these quantum handles 
is the region with nondynamical $G_{\bar{a}\bar{a}}$. In this situation 
each single quantum handle (wormhole) is the above-mentioned WH-like 
solution attached to an exterior spacetime on the surfaces where 
$u(\pm r_1) = 0$ is fulfilled.

\newpage

\end{document}